\documentclass[12pt, hidelinks]{article}
\usepackage[english]{babel}
\usepackage{amssymb,amsmath,amsthm,amsfonts,mathtools}
\usepackage[top=2.8cm, bottom=2.8cm, left=2.8cm, right=2.8cm]{geometry}
\usepackage{graphicx}
\usepackage{amsfonts}
\usepackage[utf8]{inputenc}
\usepackage{tikz}
\usepackage{tikz-cd}
\usetikzlibrary{positioning}
\usetikzlibrary{trees}
\usetikzlibrary{arrows}
\usetikzlibrary{shapes}
\usepackage{verbatim}
\usepackage{appendix}
\usepackage{hyperref}
\usepackage{booktabs}

\usepackage{datetime}
\usepackage{setspace}
\usepackage{bookmark}
\usepackage{quoting} 
\usepackage{enumerate}
\usepackage{floatrow}

\usepackage[authordate, backend=biber, natbib=true, maxcitenames=2, hyperref, backref, backrefstyle=none]{biblatex-chicago}

\DefineBibliographyStrings{english}{%
  backrefpage = {Cited on p.},
  backrefpages = {Cited on p.},
}

\usepackage{xcolor}
\usepackage{pgfplots}
\pgfplotsset{width=7cm,compat=1.8}
\usepackage{pgfplotstable}

\begin{document}

\title{Performance rating in chess, tennis, and other contexts\footnote{The author conducted this study without external funding and has no financial or non-financial conflicts of interest to disclose.}}

\author{Mehmet S. Ismail\footnote{Department of Political Economy, King's College London, London, UK. e-mail: mehmet.ismail@kcl.ac.uk}}

\date{\today}

\maketitle

\begin{abstract}
In this note, I introduce Estimated Performance Rating (PR$^e$), a novel system for evaluating player performance in sports and games. PR$^e$ addresses a key limitation of the Tournament Performance Rating (TPR) system, which is undefined for zero or perfect scores in a series of games. PR$^e$ is defined as the rating that solves an optimization problem related to scoring probability, and it is applicable for any performance level. The main theorem establishes that the PR$^e$ of a player is equivalent to the TPR whenever the latter is defined. I then apply this system to historically significant win-streaks in association football, tennis, and chess. Beyond sports, PR$^e$ has broad applicability in domains where Elo ratings are used, from college rankings to the evaluation of large language models. \textit{JEL}: D63, Z20 
\end{abstract}

\onehalfspacing

\newpage

\section{Introduction}

The practice of rating players’ performance is widespread in competitive sports and games. The Elo rating system, originally developed for chess by \citet{elo1978}, serves as a prime example. It has since been expanded to various sports, including association football, where it is used for ranking both men's and women's international teams \citep{FIFA, Hvattum2010,csato2021}.  The system is also applied in tennis \citep{Williams2021}, American football, basketball, esports, and others.  Beyond sports, the Elo-based systems are used in diverse areas, such as college rankings \citep{Avery2012}, evaluating large language models \citep{zheng2023}, dating apps \citep{Kosoff2016}, education \citep{Pelanek2016}, and biology \citep{Albers2001}.  In this system, players' ratings are dynamically updated based on their performance and the outcomes of their games.

Within this framework, the Tournament Performance Rating (TPR) is the standard method for assessing a player's performance over a series of games.  Essentially, TPR is a rating derived from the outcomes of a player's games and the Elo ratings of the opponents. To provide an intuition, suppose that a player scores $m$ points in $n$ games. Then, the player's TPR is the hypothetical rating $R$ that would remain unchanged if a player scored $m$ points in $n$ games against the same opponents. TPR is widely adopted due to its straightforward interpretability. However, it has a significant shortcoming: it is undefined in cases when a player achieves a zero or a perfect score (i.e., $m=0$ or $m=n$) in a series of games. Such cases of win- and loss-streaks are not uncommon in competitive sports. This limitation becomes particularly critical in events like tennis Grand Slams, where a player must win all matches to win the tournament. Therefore, accurately rating performances in situations of win- and loss-streaks is essential, highlighting the need for an alternative or complementary rating system to TPR in these situations.

This paper introduces a novel performance rating system, dubbed the \textit{Estimated Performance Rating} (PR$^e$), which is based on the probability of a player scoring $m\geq 0$ points in $n\geq m$ games.  Unlike the TPR, PR$^e$ is applicable to all scores. In this system, a player's win probability against a given average opponent rating is denoted as $w$, and $S(w,m,n)$ represents the probability of the player scoring $m$ points in $n$ games. The PR$^e$ is then defined as the hypothetical rating $R$ that induces  an optimal win probability $w^*$. This optimal probability is determined by solving the following maximization problem.\footnote{For a detailed definition of PR$^e$, see section~\ref{subsec:EPR}.}
\begin{equation}
\label{eq:thm}
\begin{aligned}
 \max_{w\in [0,1]} \quad & S(w,m,n) \\
\text{s.t.} \quad & S(w,m,n)\leq 0.75.\\
\end{aligned}
\end{equation}
The main theorem establishes that a rating $R$ is the TPR of the player if and only if the win probability $w$ it induces solves the maximization problem (\ref{eq:thm}) for $0<m < n$. In other words, PR$^e$ is equivalent to the TPR when $0<m < n$.

To illustrate how the PR$^e$ functions, consider a situation where a player competes in two games against opponents with an average rating of 2700. Suppose that the player achieves a score of 1.5 points, a win and a draw, from these two games. In this case, both her TPR and PR$^e$ would be calculated as 2891 (detailed analysis provided in section~\ref{subsec:example}). As mentioned above, the interpretation of her TPR is that if she had a rating of 2891, then her rating would not change after scoring 1.5 points in two games. PR$^e$, however, offers a different interpretation: if the player's rating were 2891, then the probability of her scoring exactly 1.5 points in these games would reach its maximum, calculated as 0.42 in this case. This aspect of PR$^e$ offers a unique perspective, revealing that the `predictive' value of TPR can vary (and become as low as 0.25). The maximal probability of achieving a specific score in a series of games is not constant; it varies depending on the score itself and the average ratings of the opponents (for example, see Table~\ref{tab:EPRvsTPR}).

It is worth noting that FIDE, the international governing body of chess, uses a slightly different performance rating than the TPR. This system is defined for perfect scores, but it is ad hoc and does not factor in the length of a winning or losing streak.  An illustration of the differences between TPR, FIDE's Performance Rating (FPR), and the newly proposed PR$^e$ is shown in Table~\ref{tab:EPRvsFPR}. The interpretation of PR$^e=3099$ is that if the player had a rating of 3099, then her probability of scoring 3 points in 3 games would be 0.75, meaning that the constraint in the maximization problem (\ref{eq:thm}) is binding. This interpretation will be further explored in section~\ref{subsec:EPR}, which also discusses setting different probability thresholds, other than 0.75, within the PR$^e$ framework. At the outset, predicting a rating with a 0.75 probability might not seem precise enough. However, as hinted above, the probability induced by some TPRs can be significantly lower than that.
\begin{table}[h!]
\centering
\begin{tabular}{@{}cccccccc@{}}
\toprule
$m$ & $n$  & $R_a$  & \textbf{TPR} & \textbf{FPR} &  \textbf{PR$^e$}\\ 
\midrule
 1 & 1 & 2700    & N/A & 3500 & 2891\\
  3 & 3 &2700 & N/A & 3500 & 3099\\
 5 & 5 & 2700  & N/A & 3500 & 3191\\
\bottomrule
\end{tabular}
\caption{A comparison of performance ratings: TPR, FPR, and PR$^e$}
\floatfoot{Abbreviations: $R_a$ = average rating of opponents; TPR = Tournament Performance Rating; FPR = FIDE Performance Rating; PR$^e$ =  Estimated Performance Rating; N/A: Undefined}
\label{tab:EPRvsFPR}
\end{table}

In addition to Elo-based performance rating systems mentioned above, non-Elo-based systems have also been used in economics and computer science; see, for example, \citet{guid2006}, \citet{regan2011}, \citet{kunn2022}, \citet{backus2023}, and \citet{ismail2023}. More broadly, fairness in sports has been gaining attention in recent years. Theoretical and empirical works in this area include those by \citet{Scarf2009}, \citet{apesteguia2010}, \citet{Goossens2012}, \citet{pauly2014}, \citet{kendall2017}, \citet{brams2018}, \citet{cohen2018}, \citet{brams2018b}, \citet{arlegi2020}, \citet{anbarci2021}, and \citet{Lambers2021}.\footnote{The strategy-proofness or incentive compatibility of sports rules has also been garnering more attention. Selected contributions in this area include works by \citet{brams2018b}, \citet{dagaev2018}, and \citet{csato2019,csato2021}.} For a detailed review of sports research in economics and related fields, see \citet{Palacios2023}.

\subsection{Application to tennis, association football, and chess}

In tennis, Table~\ref{tab:EPR_tennis} illustrates the Grand Slam performance ratings of 2023.
\begin{table}[h!]
\centering
\begin{tabular}{@{}llcccc@{}}
\toprule
\textbf{Player} & \textbf{Event} & $R_a$ & \textbf{Score} & \textbf{TPR} & \textbf{PR$^e$} \\ 
\midrule
Carlos Alcaraz & Wimbledon  & 1927 & 7/7  & N/A & 2478 \\ 
Novak Djokovic  & French Open  & 1867 & 7/7   & N/A & 2417 \\ 
Novak Djokovic &  Australian Open & 1865 & 7/7   & N/A & 2416 \\ 
Novak Djokovic & US Open & 1798 & 7/7   & N/A & 2349 \\ 
\bottomrule
\end{tabular}
\caption{Tennis Grand Slam performance ratings in 2023}
\label{tab:EPR_tennis}
\end{table}
Carlos Alcaraz achieves the highest PR$^e$ of 2478 with his win at Wimbledon, indicating that he won against a very strong field in this tournament (for details, see Table~\ref{tab:alcaraz}). Novak Djokovic has won three Grand Slams in 2023, and his average PR$^e$ is just under 2400.

Moving to association football, Table~\ref{tab:EPR_football} highlights the performances of teams with perfect scores in FIFA World Cups. 
\begin{table}[h!]
\centering
\begin{tabular}{@{}llcccc@{}}
\toprule
\textbf{Team} & \textbf{Event} & $R_a$ & \textbf{Score} & \textbf{TPR} & \textbf{PR$^e$} \\ 
\midrule
Brazil & Mexico 1970  & 1900 & 6/6  & N/A & 2424 \\ 
Brazil  & Korea-Japan 2002  & 1818 & 7/7   & N/A & 2369 \\ 
Italy &   France 1938 & 1802 & 4/4   & N/A & 2253 \\ 
Uruguay &  Uruguay 1930 & 1699 & 4/4   & N/A & 2150 \\ 
\bottomrule
\end{tabular}
\caption{Best performance ratings for perfect scores in FIFA World Cup history}
\label{tab:EPR_football}
\end{table}
Brazil's 1970 World Cup campaign is particularly remarkable for achieving a PR$^e$ of 2424 across six matches. This performance has long been acknowledged as one of the finest in the history of soccer. Additionally, the table illustrates other instances of perfect campaigns, including Brazil in 2002, Italy in 1938, and Uruguay in 1930.

In chess, Tables~\ref{tab:EPR_tournaments} and \ref{tab:EPR_winstreaks} present the best historical performances in tournaments and win-streaks, respectively. Bobby Fischer's 11-win streak in the 1963 USA Championship, with a PR$^e$ of 3224, stands out as an incredible achievement in chess tournament history. His 20-win streak during 1970-1971, with an even higher PR$^e$ of 3441, illustrates an unparalleled level of performance. This 20-win performance has been informally regarded as more impressive than Steinitz's 25-win streak, although this comparison had not been previously quantified in terms of TPR, as it is undefined. The tables also include notable performances by other chess legends, providing  a historical perspective on their relative achievements under a consistent performance rating system.

\begin{table}[h!]
\centering
\begin{tabular}{@{}llccccc@{}}
\toprule
\textbf{Player} & \textbf{Event} & \textbf{Year} & $R_a$ & \textbf{Score} & \textbf{TPR} & \textbf{PR$^e$} \\
\midrule
Fischer & USA Championship & 1963 & 2593 & 11/11 & N/A & 3224 \\ 
Caruana & Sinquefield Cup & 2014 & 2802 &  8.5/10 & 3103 & 3103 \\ 
Fischer &  Candidates & 1971 & 2740 & 18.5/21 & 3088 & 3088 \\ 
Alekhine & San Remo & 1930 & 2613 & 14/15 & 3072
 & 3072 \\ 
Beliavsky & Alicante & 1978 & 2392 & 13/13 & N/A & 3052 \\ 
Carlsen & Pearl Spring & 2009 & 2762 & 8/10 & 3003 & 3003 \\ 
\bottomrule
\end{tabular}
\caption{Best historical performance ratings in chess tournaments}
\label{tab:EPR_tournaments}
\end{table}

\begin{table}[h]
\centering
\begin{tabular}{@{}llcccc@{}}
\toprule
\textbf{Player} & \textbf{Event} & \textbf{Year} & $R_a$ & \textbf{Streak} & \textbf{PR$^e$} \\ 
\midrule
Fischer &  Interzonal,  Candidates & 1970--1971 & 2705 & 20-win  & 3441 \\ 
Steinitz & Vienna, London & 1873--1882 & 2581 & 25-win  & 3356 \\ 
Caruana & Sinquefield Cup & 2014 & 2793 & 7-win  & 3344 \\ 
Carlsen & Tata Steel Masters & 2015 & 2736 & 6-win  & 3260 \\ 
Fischer & USA Championship & 1963 & 2593 & 11-win  & 3224 \\ 
Carlsen & Shamkir, Grenke & 2019 & 2706 & 5-win  & 3197 \\ 
Kasparov & Wijk aan Zee & 1999 & 2632 & 7-win  & 3183 \\ 
Karpov & Linares & 1994 & 2647 & 6-win  & 3171 \\ 
Lasker & New York & 1893 & 2510 & 13-win  & 3170 \\ 
Alekhine & San Remo & 1930 & 2639 & 5-win  & 3130 \\ 
Beliavsky & Alicante & 1978 & 2392 & 13-win  & 3052 \\ 
\bottomrule
\end{tabular}
\caption{Best historical performance ratings of win-streaks in chess}
\label{tab:EPR_winstreaks}
\floatfoot{Note: For definitions of the abbreviations, see Table~\ref{tab:EPRvsFPR}.}
\end{table}

The calculation of PR$^e$ values, as presented in this note, relies on historical Elo ratings obtained from the following well-known sources: \url{www.tennisabstract.com} for tennis, \url{www.eloratings.net} for association football, and \url{www.chessmetrics.com} for chess. Detailed data supporting these calculations can be found in the Appendix. In chess, FIDE's official ratings have been used wherever applicable. In cases where a player did not have an established Elo rating, the player's TPR for the specific tournament in question has been used as a substitute. The implementation of PR$^e$, TPR, and FPR, as well as the code used to generate the values in the tables, is available at \url{www.github.com/drmehmetismail/Estimated-Performance-Rating}.

\section{Performance Ratings}

\subsection{Tournament Performance Rating}

As mentioned in the introduction, the Elo rating system serves as a standard method for ranking players based on their performance in competitive contexts, such as chess. This system assigns each player a rating. These ratings are used to calculate the probability of winning (interpreted in chess as the expected score since a draw is possible) for each player.

For two players with ratings $A$ (player 1) and $B$ (player 2), the \textbf{win probability} for player 1, denoted by $W(A,B)$, is calculated using a logistic function:
    \[
    W(A,B)=\frac{1}{1 + 10^{\frac{B - A}{400}}}.
    \]
The win probability for player 2 is simply $1 - W(A,B)$.

Note that under the Elo system, the win probability of a player with rating $A$ against a player with rating $B$ is considered independent of their win probability against a player with a different rating, say $C$. In this paper, it is also assumed that all such win probabilities are independent of each other, though this does not affect the definition of the new performance rating system I introduce. In addition, the following definitions would remain valid if one uses an extension of the Elo rating system, such as the Glicko system \citep{Glickman1995}.

A standard concept to assess player performance in a given tournament is the \textbf{Tournament Performance Rating} (TPR). To compute this, let $b_1, b_2, ..., b_k$ represent a sequence of ratings of opponents faced by player 1. The \textbf{average rating} of these opponents, $R_a$, is given by:
    \[
    R_a=\frac{1}{k}\sum^{k}_{j=1} b_j.
    \]
Suppose that player 1 scores $m\geq 0$ points in a total of $n>m$ games against these opponents. The TPR is defined by the equation:
    \[
    m=\frac{n}{1 + 10^{\frac{R_a - TPR}{400}}}.
    \]
This can be further expressed as:
\begin{equation}
\label{eq:TPR}
        \frac{m}{n}=\frac{1}{1 + 10^{\frac{R_a - TPR}{400}}}.
\end{equation}

\subsection{FIDE's Performance Rating}

The performance rating system used by FIDE slightly differs from the TPR. The \textbf{FIDE Performance Rating} (FPR) is calculated by adding a rating difference ($dp$), which is based on the percentage score, to the average rating of opponents ($R_a$).  Although FPR does not exactly match with the TPR, the results are generally similar. Importantly, FPR plays a crucial role in determining ``norms,'' which are sets of criteria required to achieve titles such as Grandmaster (GM) and International Master (IM). FPR is defined as follows:
\[
FPR = R_a + dp,
\]
where the rating difference $dp$ is determined by the player's score percentage ($ps$) as outlined in Table~\ref{tab:fide} \citep{FIDE2022}. 

It is important to note that for a perfect score, whether it be 7/7 or 11/11, FIDE assigns a $dp$ value of 800 as illustrated in Table~\ref{tab:EPRvsFPR}. FIDE recognizes that for a zero or perfect score ``dp is necessarily indeterminate but is shown \textit{notionally} as 800'' (emphasis added). Historically, these calculations were manually performed by FIDE officials, which is one of the reasons why FIDE uses a predefined table rather than the original TPR formula.

\subsection{Estimated Performance Rating}
\label{subsec:EPR}
Assume that player 1 scores $m$ points in $n$ games against players with an average rating $R_a$, where $m$ is an integer.\footnote{If $m$ is not integer then multiply, without loss of generality, both $m$ and $n$ by 2 to make $m$ integer. Recall that in chess, a win is worth 1 point, a draw 0.5 points, and a loss 0 points in chess.} Let $S(w,m,n)$ denote the player 1's \textbf{probability of scoring} exactly $m$ points in $n$ games, given player 1's win probability $w$ against players with an average rating $R_a$. 
\[
S(w,m,n) = {n\choose m} w^m (1-w)^{n-m}.
\]
Similarly, let $\bar{S}(w,m,n)$ denote the probability of player 1 scoring  $m$ points or more in $n$ games, given $w$. 
\[
\bar{S}(w,m,n) = \sum^{n}_{k=m} {n\choose k} w^k (1-w)^{n-k}.
\]

For a given threshold $t\in [0,1]$, define the following maximization problem to find $w$ that maximizes $S(w,m,n)$.\footnote{Alternatively,  $\bar{S}(w,m,n)$ can be used in this maximization problem.}
\begin{equation}
\label{eq:max}
\begin{aligned}
 \max_{w\in [0,1]} \quad & S(w,m,n) \\
\text{s.t.} \quad & S(w,m,n)\leq t.\\
\end{aligned}
\end{equation}

Unless otherwise noted in this paper, I set $t=0.75$, indicating that for a given $w$, it is 75\% likely that player 1 scores $S(w,m,n)$.

Let $w^*$ be the value that solves the optimization problem (\ref{eq:max}). Then, given $R_a$, find the value $A^*$ such that 
\begin{equation}
\label{eq:EPR}
  W(A^*,R_a)=w^*=\frac{1}{1 + 10^{\frac{R_a - A^*}{400}}}.
\end{equation}

In this context, $A^*$ is called the \textbf{Estimated Performance Rating} (PR$^e$) of player 1, given the score $m$ in $n$ games and the average rating $R_a$ of the opposition. Here, $PR^e(w^*,R_a)$ denotes the performance rating of player 1 given $w^*$ and $R_a$. Note that $w^*$ is dependent on $t$, $m$, and $n$.

The next step involves solving Equation \ref{eq:EPR} for  $A^*$.
Begin by cross-multiplying to obtain:
    \[ w^* + w^* \cdot 10^{\frac{R_a - A^*}{400}} = 1 \]
Next, we proceed to rearrange the terms:
$10^{\frac{R_a - A^*}{400}} = \frac{1 - w^*}{w^*}$. 
Applying the logarithm to both sides, we get:
    \[ \frac{R_a - A^*}{400} = \log_{10}\left( \frac{1 - w^*}{w^*} \right). \]
Rearranging the equation yields the solution for  $A^*$, as shown in the equation below:
    \begin{equation}
    \label{eq:EPR2}
    A^* = R_a - 400 \cdot \log_{10}\left( \frac{1 - w^*}{w^*} \right).
    \end{equation}

\subsection{Illustrative Example}
\label{subsec:example}

\begin{table}[h]
\centering
\begin{tabular}{@{}cccccccc@{}}
\toprule
$R_a$ & $m$ & $n$  & $w^*$ & $S(w^*,m,n)$ & \textbf{PR$^e$} & \textbf{TPR} \\ 
\midrule
2700 & 0 & 2    & 0.13 & 0.75 & 2376 & N/A \\
2700 & 0.5 & 2  & 0.250 & 0.42   & 2509 & 2509 \\
2700 & 1 & 2    & 0.50 & 0.50   & 2700 & 2700 \\
2700 & 1.5 & 2  & 0.75 & 0.42  & 2891 & 2891 \\
2700 & 2 & 2   & 0.87 & 0.75  & 3024 & N/A \\
\bottomrule
\end{tabular}
\caption{Illustrative example of performance ratings based on different scores}
\label{tab:EPRvsTPR}
\end{table}

To illustrate how TPR and PR$^e$ are calculated, consider the example in Table~\ref{tab:EPRvsTPR}. In this example, player 1 has an average rating of 2700 and plays 2 games against players with an average rating of 2700. 
$S(w^*,m,n)$ shows the probability of scoring $m$ points in $n$ games given $w^*$, which is derived from the optimization problem (\ref{eq:max}).

For the given score $m = 1$ and $n = 2$, I calculate the TPR and PR$^e$. For TPR, I use the formula in Equation~\ref{eq:TPR} and for PR$^e$, I use the formula in Equation~\ref{eq:EPR}. 
Solving the following equation for TPR
\[
\frac{1}{2}=\frac{1}{1 + 10^{\frac{2700 - TPR}{400}}},
\]
yields TPR $=2700$. Now, I calculate the PR$^e$. For $m=1$ and $n=2$, we have $w^*=0.5$. Then, plugging 
$w^*=0.5$ and $R_a=2700$ into the formula for PR$^e$, we obtain $PR^e=2700$.

Finally, I calculate PR$^e$ for $m=0$ and $n=2$. (Note that TPR is undefined for $m=0$ and $m=2$.) 
For $m=0$ and $n=2$, solving the optimization problem (\ref{eq:max}) yields $w^*=0.29$. Then, plugging 
$w^*=0.29$ and $R_a=2700$ into the formula for PR$^e$, we obtain PR$^e$ $=2546.89$. The remainig values of 
PR$^e$ and TPR are calculated similarly.

\section{Main Result}
\label{sec:thm}

\begin{figure}[ht]
    \centering
    \includegraphics[width=0.45\textwidth]{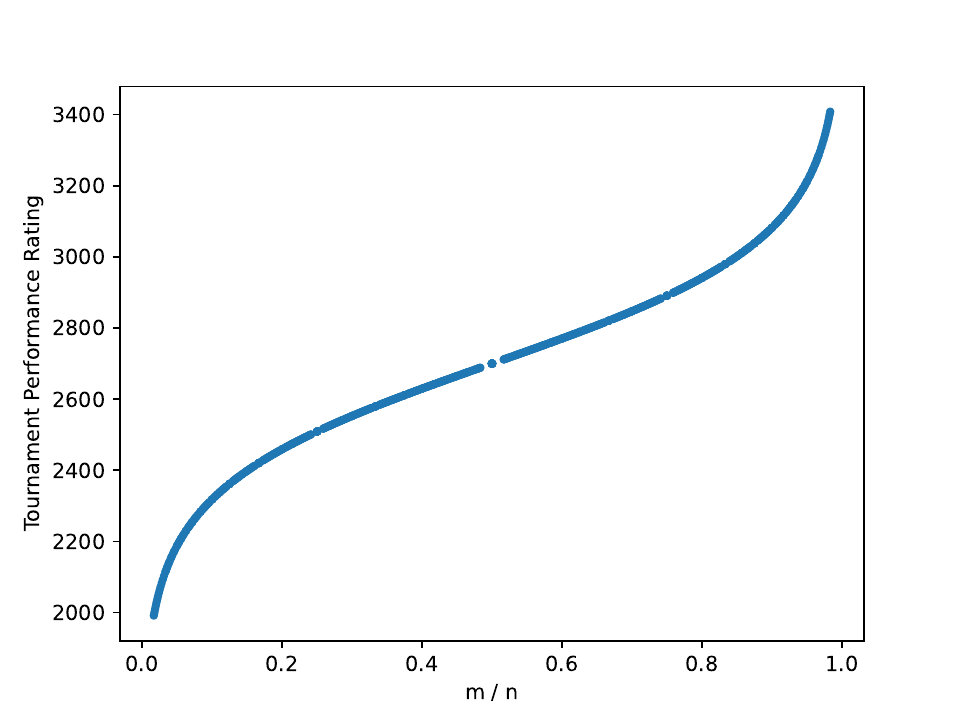}
    \hfill
    \includegraphics[width=0.45\textwidth]{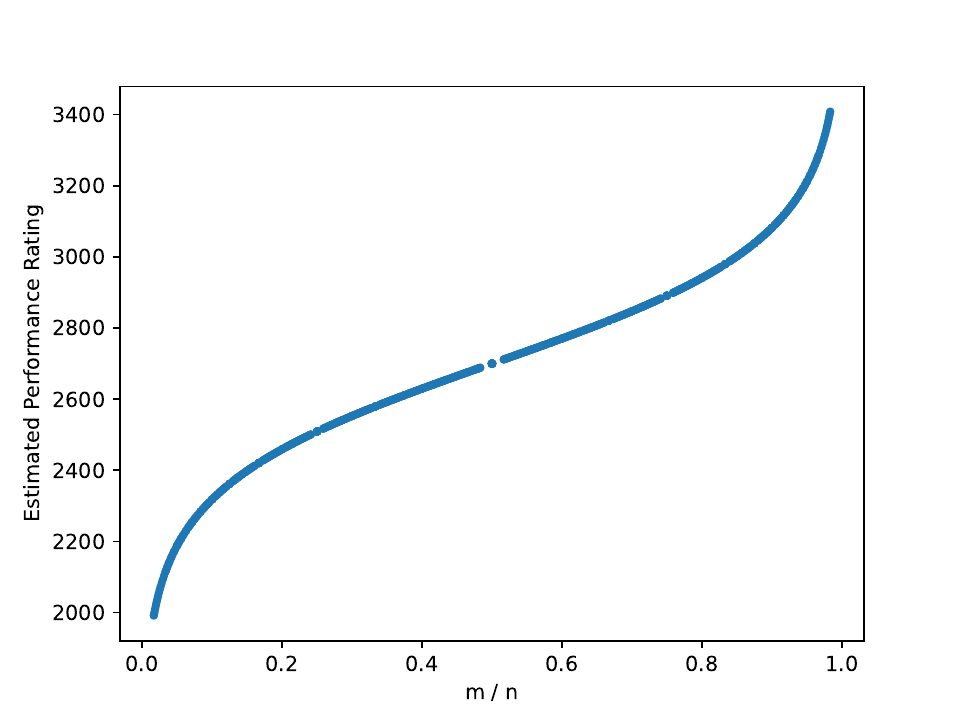}
    \caption{Plots of TPR and PR$^e$ for every $m$ and $n \leq 30$, where $0<m<n$.}
    \label{fig:thm}
\end{figure}

Figure~\ref{fig:thm} illustrates that TPR and PR$^e$ coincide for every $m$ and $n\leq 30$ such that $0<m<n$. The main result establishes that this pattern holds whenever $0<m<n$.

\bigskip\noindent
\textbf{Main Theorem.}
\textit{
    Let $m$ be the score of a player in $n$ games such that $0<m<n$. The rating $R$ is the TPR of the player if and only if $W(R, R_a) \in \arg\max_{w \in [0, 1]} S(w, m, n)$.}

\begin{proof}
Consider the function to be maximized:
\begin{equation*}
    \max_{w \in [0, 1]} \quad \binom{n}{m} w^m (1 - w)^{n - m}
\end{equation*}
where $m$ and $n$ are constants, and $w\in [0,1]$.

Define $f(w) = \binom{n}{m} w^m (1 - w)^{n - m}$. Taking the derivative of $f$ with respect to $w$, we obtain:
\begin{equation*}
    f'(w) = \binom{n}{m} \left[ m w^{m-1} (1 - w)^{n-m} - w^m (n - m) (1 - w)^{n-m-1} \right].
\end{equation*}
To identify the critical points, set the derivative to zero:
\begin{equation*}
    f'(w) = 0.
\end{equation*}
This leads to the equation:
\begin{equation*}
    \binom{n}{m} \left[ m w^{m-1} (1 - w)^{n-m} - w^m (n - m) (1 - w)^{n-m-1} \right] = 0
\end{equation*}
Simplifying the equation, we obtain:
\begin{equation*}
    m w^{m-1} (1 - w)^{n-m} = w^m (n - m) (1 - w)^{n-m-1}.
\end{equation*}
Dividing both sides by $w^{m-1} (1 - w)^{n-m-1}$ yields:
\begin{equation*}
    m (1 - w) = w (n - m).
\end{equation*}
By rearranging, we find the critical value:
\begin{equation*}
    w^* = \frac{m}{n}.
\end{equation*}
Next, evaluate the second derivative of $f(w)$ at $w = \frac{m}{n}$: 
\[
f''(\frac{m}{n}) = \binom{n}{m} \frac{n^3 (\frac{m}{n})^m \left(1 - \frac{m}{n}\right)^{n-m}}{m(m - n)}.
\]
Since $m < n$, $f''(w)$ is negative at this point. Therefore, $w^* = \frac{m}{n}$ maximizes $f(w)$.

Next, assuming $R$ is the TPR, by Equation~\ref{eq:TPR}, we have:
\begin{equation*}
        \frac{m}{n} = \frac{1}{1 + 10^{\frac{R_a - R}{400}}}.
\end{equation*}
This holds if and only if
\begin{equation*}
  W(R, R_a) = w^* = \frac{1}{1 + 10^{\frac{R_a - R}{400}}}.
\end{equation*}
Thus, $W(R, R_a) = w^*$ is a solution to the optimization problem (\ref{eq:max}) when $0<m<n$.
\end{proof}

\begin{figure}
    \centering
    \includegraphics[width=0.7\textwidth]{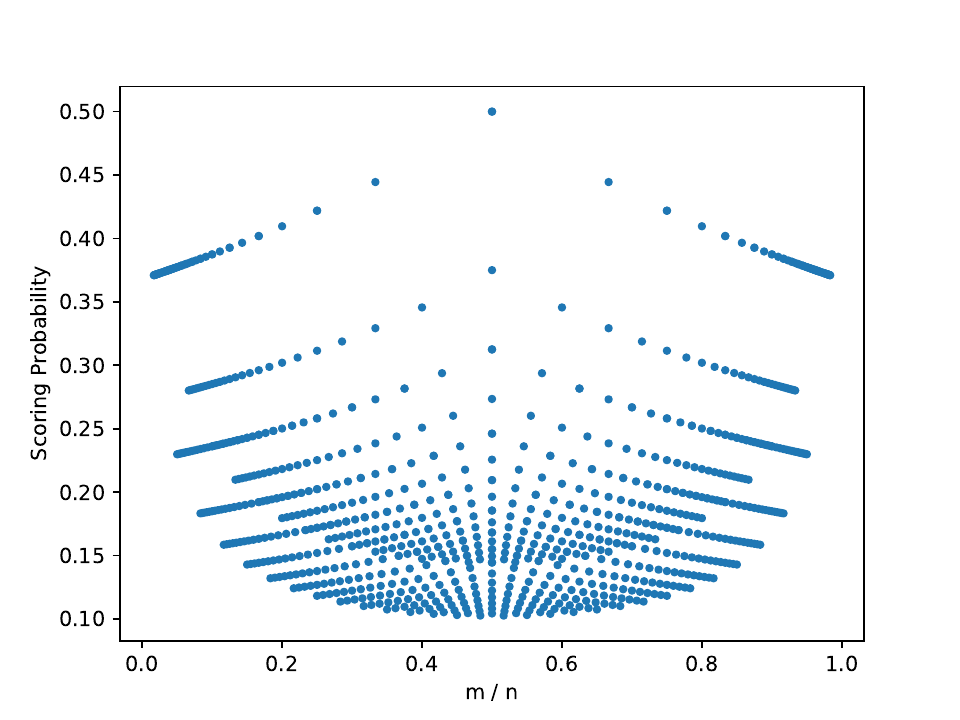}
    \caption{Plot of $f(\frac{m}{n})$ for every $m$ and $n \leq 30$.}
    \label{fig:plot}
\end{figure}

It is instructive to examine the behavior of the score probability function at its maximum, $S(\frac{m}{n}, m, n)$. The value of the function at $w = \frac{m}{n}$ is:
\[
f(\frac{m}{n}) = (\frac{m}{n})^m (1 - \frac{m}{n})^{n-m}.
\]
Figure~\ref{fig:plot} illustrates the value of $f(w)$ for various values of $w=\frac{m}{n}$. Observe that the function reaches its maximum when $\frac{m}{n} = 0.5$, particularly when $m = 0.5$ and $n = 1$. This is intuitive since, for larger values of $n$, $m$ represents just one among many possible scores less than or equal to $n$. 

\printbibliography

\section*{Appendix}

\begin{table}[h]
\centering
\begin{tabular}{@{}llc@{}}
\toprule
\textbf{Player} & \textbf{Tournament} & \textbf{Elo} \\ \midrule
Baena & AUS Open & 1742 \\
Couacaud & AUS Open & 1564 \\
Dimitrov & AUS Open & 1888 \\
de Minaur & AUS Open & 1945 \\
Rublev & AUS Open & 1970 \\
Paul & AUS Open & 1886 \\
Tsitsipas & AUS Open & 2058 \\
Kovacevic & French Open & 1669 \\
Fucsovics & French Open & 1783 \\
Fokina & French Open & 1864 \\
Varillas & French Open & 1687 \\
Khachanov & French Open & 1960 \\
Alcaraz & French Open & 2190 \\
Ruud & French Open & 1918 \\
Muller & US Open & 1658 \\
Miralles & US Open & 1750 \\
Djere & US Open & 1812 \\
Gojo & US Open & 1659 \\
Fritz & US Open & 1961 \\
Shelton & US Open & 1643 \\
Medvedev & US Open & 2101 \\
\bottomrule
\end{tabular}
\caption{Djokovic's Opponents in 2023 Grand Slams}
\label{tab:Djokovic}
\end{table}

\begin{table}[h]
\centering
\begin{tabular}{@{}llc@{}}
\toprule
\textbf{Player} & \textbf{Tournament} & \textbf{Elo} \\ \midrule
Chardy & Wimbledon & 1808 \\
Berrettini & Wimbledon & 1848 \\
Muller & Wimbledon & 1660 \\
Jarry & Wimbledon & 1839 \\
Medvedev & Wimbledon & 2110 \\
Rune & Wimbledon & 2050 \\
Djokovic & Wimbledon & 2171 \\
\bottomrule
\end{tabular}
\caption{Alcaraz's opponents in Wimbledon 2023}
\floatfoot{Note: Chardy currently does not have an Elo rating; therefore, his last available Elo rating was used.}
\label{tab:alcaraz}
\end{table}

\begin{table}[h]
\centering
\begin{tabular}{@{}llcc@{}}
\toprule
\textbf{Opponent} & \textbf{Score} & \textbf{Rating}  \\ \midrule
Argentina & 4-2 & 2084 \\
Yugoslavia & 6-1 & 1608 \\
Romania & 4-0 & 1560  \\
Peru & 1-0 & 1542  \\
\bottomrule
\end{tabular}
\caption{Uruguay's Matches in 1930}
\end{table}

\begin{table}[h]
\centering
\begin{tabular}{@{}llcc@{}}
\toprule
\textbf{Opponent} & \textbf{Score} & \textbf{Rating}  \\ \midrule
Hungary & 4-2 & 1953 \\
Brazil & 2-1 & 1908 \\
France & 3-1 & 1618 \\
Norway & 2-1 & 1729 \\
\bottomrule
\end{tabular}
\caption{France's Matches in 1938}
\end{table}

\begin{table}[h]
\centering
\begin{tabular}{@{}llcc@{}}
\toprule
\textbf{Opponent} & \textbf{Score} & \textbf{Rating}  \\ \midrule
Italy & 4-1 & 2004 \\
Uruguay & 3-1 & 1863 \\
Peru & 4-2 & 1707 \\
England & 1-0 & 2087 \\
Romania & 3-2 & 1791 \\
Czechoslovakia & 4-1 & 1947 \\
\bottomrule
\end{tabular}
\caption{Brazil's Matches in 1970}
\end{table}

\begin{table}[h]
\centering
\begin{tabular}{@{}llc@{}}
\toprule
\textbf{Opponent} & \textbf{Score} & \textbf{Rating}  \\ \midrule
Germany & 2-0 & 1869 \\
Turkey & 1-0 & 1797 \\
England & 2-1 & 1932 \\
Belgium & 2-0 & 1835 \\
China & 4-0 & 1726 \\
Costa Rica & 5-2 & 1772 \\
Turkey & 2-1 & 1797 \\
\bottomrule
\end{tabular}
\caption{Brazil's Matches in Korea-Japan 2002}
\end{table}

\begin{table}[h]
\centering
\begin{tabular}{@{}llcc@{}}
\toprule
\textbf{Opponent} & \textbf{Score} & \textbf{Rating} & \textbf{Tournament} \\ \midrule
Rosenthal  & 2-0 & 2571 & Vienna, 1873 \\
Paulsen & 2-0 & 2624 & Vienna, 1873 \\
Anderssen & 2-0 & 2648 & Vienna, 1873 \\
Schwarz  & 2-0 & 2481 & Vienna, 1873 \\
Gelbfuhs & 2-0 & 2439 & Vienna, 1873 \\
Bird & 2-0 & 2589 & Vienna, 1873 \\
Heral & 2-0 & 2487 & Vienna, 1873 \\
Blackburne & 2-0 & 2578 & Vienna, 1873 \\
Blackburne & 7-0  & 2648 & London, 1876 \\
Blackburne & 1-0 & 2716 & Vienna, 1882 \\
Noa & 1-0 & 2449 & Vienna, 1882 \\ 
\bottomrule
\end{tabular}
\caption{Steinitz's games in Vienna, 1873 and 1882, and in London 1876}
\end{table}

\begin{table}[h]
\centering
\begin{tabular}{@{}lcc@{}c}
\toprule
\textbf{Opponent} & \textbf{Rating} & \textbf{Score} & \textbf{Tournament}\\ \midrule
Jorge Alberto Rubinetti & 2503  & 1 - 0    & Interzonal, 1970                            \\
Wolfgang Uhlmann  & 2685     & 1 - 0        & Interzonal, 1970                           \\
Mark E Taimanov   & 2731   & 1 - 0          & Interzonal, 1970                            \\
Duncan Suttles    & 2581   & 1 - 0          & Interzonal, 1970                            \\
Henrique Mecking  & 2619   & 1 - 0          & Interzonal, 1970                            \\
Svetozar Gligoric & 2693   & 1 - 0          & Interzonal, 1970                            \\
Oscar Panno       & 2583   & 1 - 0          & Interzona, 1970l   \\
Mark Taimanov     & 2731            & 6 - 0          & Candidates, 1971                      \\
Bent Larsen       & 2752            & 6 - 0   & Candidates, 1971                             \\
Tigran Petrosian  & 2738            & 1 - 0  & Candidates, 1971                         \\ \bottomrule
\end{tabular}
\caption{Fischer's 20-game win streak}
\end{table}

\begin{table}[h]
\centering
\begin{tabular}{@{}lccc@{}}
\toprule
\textbf{Opponent} & \textbf{Rating} & \textbf{Score} & \textbf{Tournament} \\ \midrule
Magnus Carlsen        & 2877 & 1 - 0 & Sinquefield Cup, 2014 \\
Veselin Topalov       & 2772 & 2 - 0 & Sinquefield Cup, 2014 \\
Maxime Vachier-Lagrave & 2768 & 2 - 0 & Sinquefield Cup, 2014 \\
Levon Aronian         & 2805 & 1 - 0 & Sinquefield Cup, 2014 \\
Hikaru Nakamura       & 2787 & 1 - 0 & Sinquefield Cup, 2014 \\
\bottomrule
\end{tabular}
\caption{Caruana's 7-Game win streak at the Sinquefield Cup 2014}
\end{table}

\begin{table}
\begin{center}
\begin{tabular}{|c|c||c|c||c|c||c|c||c|c||c|c|}
\hline
ps & dp & ps & dp & ps & dp & ps & dp & ps & dp & ps & dp \\
\hline
1.0 & 800 & .83 & 273 & .66 & 117 & .49 & -7 & .32 & -133 & .15 & -296 \\
.99 & 677 & .82 & 262 & .65 & 110 & .48 & -14 & .31 & -141 & .14 & -309 \\
.98 & 589 & .81 & 251 & .64 & 102 & .47 & -21 & .30 & -149 & .13 & -322 \\
.97 & 538 & .80 & 240 & .63 & 95 & .46 & -29 & .29 & -158 & .12 & -336 \\
.96 & 501 & .79 & 230 & .62 & 87 & .45 & -36 & .28 & -166 & .11 & -351 \\
.95 & 470 & .78 & 220 & .61 & 80 & .44 & -43 & .27 & -175 & .10 & -366 \\
.94 & 444 & .77 & 211 & .60 & 72 & .43 & -50 & .26 & -184 & .09 & -383 \\
.93 & 422 & .76 & 202 & .59 & 65 & .42 & -57 & .25 & -193 & .08 & -401 \\
.92 & 401 & .75 & 193 & .58 & 57 & .41 & -65 & .24 & -202 & .07 & -422 \\
.91 & 383 & .74 & 184 & .57 & 50 & .40 & -72 & .23 & -211 & .06 & -444 \\
.90 & 366 & .73 & 175 & .56 & 43 & .39 & -80 & .22 & -220 & .05 & -470 \\
.89 & 351 & .72 & 166 & .55 & 36 & .38 & -87 & .21 & -230 & .04 & -501 \\
.88 & 336 & .71 & 158 & .54 & 29 & .37 & -95 & .20 & -240 & .03 & -538 \\
.87 & 322 & .70 & 149 & .53 & 21 & .36 & -102 & .19 & -251 & .02 & -589 \\
.86 & 309 & .69 & 141 & .52 & 14 & .35 & -110 & .18 & -262 & .01 & -677 \\
.85 & 296 & .68 & 133 & .51 & 7 & .34 & -117 & .17 & -273 & .00 & -800 \\
.84 & 284 & .67 & 125 & .50 & 0 & .33 & -125 & .16 & -284 & & \\
\hline
\end{tabular}
\caption{FIDE's predefined table for the calculation of  the rating difference (dp) based on percentage score (ps)}
\label{tab:fide}
\end{center}
\end{table}

\end{document}